\documentclass[twocolumn,aps,showpacs,prb,tightenlines,amsmath,amssymb,superscriptaddress]{revtex4}
\usepackage{graphicx}
\usepackage{amssymb}
\usepackage{dcolumn}
\usepackage{amsmath}
\usepackage{bm}

\usepackage{colordvi}
\newcommand{\bgreek}[1]{\mbox{\boldmath$#1$\unboldmath}}

\begin{document}

\title{Multivalley spin relaxation in the presence of high in-plane
 electric fields
in $n$-type GaAs quantum wells}

\author{P. Zhang}
\affiliation{Hefei National Laboratory for Physical Sciences at
Microscale,
University of Science and Technology of China, Hefei,
Anhui, 230026, China}
\affiliation{Department of Physics,
University of Science and Technology of China, Hefei,
Anhui, 230026, China}
\author{J. Zhou}
\affiliation{Department of Physics,
University of Science and Technology of China, Hefei,
Anhui, 230026, China}
\altaffiliation{Mailing address}
\author{M.\ W.\ Wu}
\thanks{Author to  whom correspondence should be addressed}
\email{mwwu@ustc.edu.cn.}
\affiliation{Hefei National Laboratory for Physical Sciences at
Microscale,
University of Science and Technology of China, Hefei,
Anhui, 230026, China}
\affiliation{Department of Physics,
University of Science and Technology of China, Hefei,
Anhui, 230026, China}
\altaffiliation{Mailing address}

\date{\today}

\begin{abstract}
Multi-valley spin relaxation in $n$-type GaAs quantum wells
with in-plane electric field is investigated at high temperature
by means of kinetic spin
Bloch equation approach. The spin relaxation time first increases and
then decreases with electric
field, especially when the temperature is relatively low. We show that
$L$ valleys
play the role of a ``drain'' of the total spin polarization due to the
large spin-orbit coupling in $L$ valleys and the strong $\Gamma$-$L$
inter-valley scattering, and thus can enhance spin relaxation of the
total system effectively when the in-plane electric field is high.
Under electric field,  spin
precession resulting from the electric-field-induced
magnetic field is observed. Meanwhile, due to the strong $\Gamma$-$L$
inter-valley scattering as well as the
strong inhomogeneous broadening in $L$ valleys,
electron spins in $L$ valleys possess almost the same damping rate and
precession frequency as those in $\Gamma$ valley.
This feature still holds when a finite static magnetic
field is applied in Voigt configuration, despite that
the  $g$-factor of $L$ valleys is much larger than that of
$\Gamma$ valley. Moreover, it is shown that the property of  spin precession
of the whole system is dominated by electrons in $\Gamma$ valley.
Temperature, magnetic field, and
impurity can affect spin relaxation in low electric field regime.
However, they are shown to have marginal influence in high electric field regime.
\end{abstract}

\pacs{72.25.Rb, 72.20.Ht, 71.10.-w, 67.57.Lm}

\maketitle

\section{Introduction}
The ability to manipulate electron spin degree of freedom in
semiconductors has become an important issue for the realization of
spintronic devices.\cite{meier,aws,zutic} Usual ways to manipulate spins
are proposed by directly applying magnetic field or exploiting the spin-orbit
interaction through a gate voltage\cite{bychkov1,bychkov2,nitta} and/or strain
field.\cite{kato,strain,crooker1,beck}
In addition, applying a drift electric field also
provides an approach to manipulate spin, by which the dependence
of spin-orbit coupling on electron momentum is exploited and an
effective magnetic field can be induced.\cite{weng1}
However, these manipulations also affect spin
relaxation and dephasing
(R\&D) time,\cite{pramanik1,pramanik2,barry,weng1,beck} which is an
essential time scale for the design of spin-based devices.
Moreover, as the goal of semiconductor spintronics is
to combine ``traditional'' semiconductor electronics with
the utilization of the spin state and most current electronic devices are
performed at high-field transport condition,
the study of drift electric field on spin R\&D is therefore
essential.

Earlier Monte-Carlo simulation has revealed that the drift electric
field can enhance spin relaxation in GaAs quantum
wires\cite{pramanik1,pramanik2} and bulk material.\cite{barry} However,
these studies fail to treat all scattering,
especially the Coulomb scattering, completely. It was
pointed out that the Coulomb scattering can also cause
spin R\&D.\cite{wu2,ivc}
Investigations based on fully microscopic kinetic spin Bloch
equation (KSBE) approach\cite{wu1,wu-rev} have demonstrated that the Coulomb
scattering is very important to the spin R\&D.\cite{wu2,weng1,weng3,weng2,zhou}
This has also been verified experimentally.\cite{harley,tob1,tob2}
Among these works, Ref.\ \onlinecite{weng1} gives a complete
understanding of the hot-electron effect on spin
R\&D in $n$-type GaAs quantum wells (QWs),
where all the scattering is explicitly included and calculated
self-consistently. It is shown that the spin R\&D increase
with electric field at high temperature,
 but decrease with it at low temperature.\cite{weng1}
Nevertheless, due to the
lowest-subband used in the investigation and the ``runaway effect'',\cite{dmitriev} that
study is limited to electric field lower than 1\ kV/cm.\cite{weng1}
When the electric field is further raised, the multi-subband or even
the multi-valley effect should be taken into
account.\cite{weng2} In Ref.\ \onlinecite{weng2}, the electric field has been
raised up to 3\ kV/cm and the multi-subband scattering is considered.
It is shown that the spin R\&D first decrease then increase with the
applied electric field at low temperature.\cite{weng2} These
effects\cite{weng2,strain} have very recently been demonstrated
by  Holleitner {\em et al.} experimentally.\cite{david}
However, when the electric field is
further increased,  the inter-valley scattering
should be taken into account.

The inter-valley scattering can lead to interesting phenomena, such as
the charge Gunn effect in GaAs,\cite{gunn}
which comes from the different mobilities
in different valleys due to the different effective masses. Studies focusing
on scattering and transport between different valleys without
the spin degree of freedom have been performed
long time ago.\cite{lei,xing,magnus,educato,sjakste} When
spin is considered, the different spin-orbit couplings and the
different effective $g$-factors, together with the well known
different momentum relaxations at different valleys
may lead to complex  spin R\&D phenomena. Spin injection
through a Schottky barrier into bulk GaAs, with $\Gamma$-$L$-$X$ valley
structure included, has been studied by Monte-Carlo
simulation, revealing faster decay of current (electron) spin
polarization in the upper valleys.\cite{saikin} Recently, a spin Gunn
effect has been proposed, depicting a process of spontaneous
spin-polarization amplification in semiconductors such as
GaAs.\cite{flatte} However, as this effect is
predicted to happen in the high-electric-field charge Gunn domain, and
the condition for it depends strongly on spin
relaxation time, it is necessary to investigate spin
relaxation under high electric field.

In this work, we perform a fully microscopic investigation on spin
relaxation with high in-plane electric
field at high temperature in $n$-type GaAs QWs, with
 $\Gamma$ and $L$ valleys included. The spin
relaxation comes from the D'yakonov-Perel' mechanism.\cite{dyakonov}
This paper is organized as follows. In Sec.\ II, we set up the model
and construct the KSBEs. We present our main
results in Sec.\ III, including the role of inter-valley
scattering, the spin precession properties,
and the effects of temperature, magnetic field, and
impurity on spin R\&D. We summarize in Sec.\ IV.

\section{Model and KSBEs}
We start our investigation from
an $n$-type GaAs [001] QW with well width $a$. The growth direction is assumed
along $z$-axis.
 $\Gamma$ valley is located at the
center of the Brillouin zone
and four $L$ valleys are  at
${\cal \bf K}^{0}_{L_{i}}$=$(\pi/a_{0})(1,\pm 1,\pm 1)$
that are rotationally symmetric around the
[100] ($x$) axis, with $a_{0}$ denoting the lattice constant and $i=1$,$\cdots$,4.
Under spherically symmetric approximation, the effective
electron masses of $\Gamma$ and $L$ valleys are
$m_{\Gamma}$=0.067m$_{0}$ and $m_{L_{i}}$=$m_{L}$=0.230$m_{0}$,\cite{lei,chand} where
$m_{0}$ is the free-electron mass. Assuming parabolic and isotropic
band structure for each valley, the electron energy at the
bottom of each valley reads\cite{magnus}
\begin{eqnarray}
&&\varepsilon_{{\bf k}_{\Gamma}}=\frac{\hslash^{2}{\bf k}_{\Gamma}^{2}}{2m_{\Gamma}}\ ,
\\
&&\varepsilon_{{\bf k}_{L_{i}}}=\frac{\hslash^{2}{\bf
      k}_{L_{i}}^{2}}{2m_{L}}+\varepsilon_{\Gamma L}\ ,
\end{eqnarray}
where ${\bf k}_{\lambda}$ is the $x$-$y$-plane projection of the
three-dimensional relative momentum ${\bf K}-{\bf
  K}_{\lambda}^{0}$ ($\lambda$=$\Gamma$, $L_{i}$), with the
electron energy at ${\bf K}^{0}_{\Gamma}$  set as the reference
point. $\varepsilon_{\Gamma L}$ is the effective $\Gamma$-$L$
energy difference in the two-dimensional structure:
$\varepsilon_{\Gamma L}=E_{\Gamma L}-\frac{\hslash^{2}\pi^{2}}
{2a^{2}}(\frac{1}{m_{\Gamma}}-\frac{1}{m_{L}}$),
where $E_{\Gamma L}$=0.28\ eV is the $\Gamma$-$L$ energy difference in
bulk.\cite{pozela} Here an infinite-well-depth assumption
is used. $a$ is set as 7.5\ nm throughout
the paper.  $\varepsilon_{\Gamma L}=0.21$\ eV is smaller than
the energy difference of the lowest two subbands of the $\Gamma$ valley,
which is about 0.30\ eV. It is also noted that the effective $L$-$X$
energy difference in this two dimensional structure is as high as 0.17\
eV when the spherical effective mass of $X$ valleys is about
$0.6m_{0}$.\cite{chand} Thus in this work, only the lowest
subbands of $\Gamma$ and $L$ valleys  are considered.

The Hamiltonian of the system reads
\begin{eqnarray}
&&H=\sum_{\lambda}H^{0}_{\lambda}+\sum_{\lambda\lambda^{\prime}}H^{I}_{\lambda\lambda^{\prime}}\ ,\\
&&H^{I}_{\lambda\lambda^{\prime}}=H_{\lambda\lambda^{\prime}}^{e-e}+H_{\lambda\lambda^{\prime}}^{e-p}
+H_{\lambda\lambda^{\prime}}^{e-i}\delta_{\lambda\lambda^{\prime}}\ .
\end{eqnarray}
Here
\begin{eqnarray}
\nonumber
H_{\lambda}^{0}&=&\sum_{{{\bf k}_{\lambda}}\sigma\sigma^{\prime}}\Big\{(\varepsilon_{{\bf
      k}_{\lambda}}+e{\bf E}\cdot{\bf R})\delta_{\sigma\sigma^{\prime}}\\
&& \mbox{}+[\frac{1}{2}g_{\lambda}\mu_{B}{\bf B}+{\bf h}_{\lambda}({\bf k}_{\lambda})]
\cdot{\bgreek \sigma}_{\sigma\sigma^{\prime}}\Big\}C_{{\bf k}_{\lambda}\sigma}^{\dagger}
C_{{\bf k}_{\lambda}\sigma^{\prime}}
\end{eqnarray}
is the free electron Hamiltonian at $\lambda$ valley, where
 $-e$ is the electron charge, ${\bf R}$=($x$, $y$) is
the position of the electron, ${\bf E}$ is the electric field (the
system is assumed to be spacial homogeneous one
so that ${\bf E}$ does not depend on
the position), ${\bf B}$ is the magnetic field in Voigt configuration
and ${\bgreek \sigma}$ is
Pauli matrices. In the calculation, ${\bf E}$ and
${\bf B}$ are taken along $x$-axis unless
  otherwise specified.
$C_{{\bf k}_{\lambda}\sigma}^{\dagger}$($C_{{\bf k}_{\lambda}\sigma}$) is the
creation (annihilation) operator for
electron with relative wave vector ${\bf
  k}_{\lambda}$ and spin ${\sigma}$. The effective $g$-factors are
$g_{\Gamma}=-0.44$\cite{land} and $g_{L_{i}}=g_{L}=1.77$.\cite{shen}
${\bf h}_{\lambda}({\bf k})$
represents the Dresselhaus spin-orbit coupling.\cite{dres}
 In the  coordinate system adopted in this work,\cite{ivchenko}
\begin{eqnarray}
h_{\Gamma}({\bf k}_{\Gamma})&=&\frac{\gamma}{2}\big(k_{\Gamma x}(k_{\Gamma y}^{2}-\langle
k_{\Gamma z}^{2}\rangle),k_{\Gamma y}(\langle k_{\Gamma
  z}^{2}\rangle-k_{\Gamma x}^{2}), \nonumber\\
&&\langle k_{\Gamma z}\rangle(k_{\Gamma
  x}^{2}-k_{\Gamma y}^{2})\big)\ ,
\label{hg}\\
h_{L_{i}}({\bf
  k}_{L_{i}})&=&\beta\big(k_{L_{i}x},k_{L_{i}y},\langle
k_{L_{i}z}\rangle\big)\times\hat{{\bf n}}_{i}\ .
\label{eqso}
\end{eqnarray}
Here $\hat{\bf n}_{i}$ is the unit vector along the longitudinal principle
axis of $L_{i}$ valley. $\langle k_{\lambda z}\rangle$ ($\langle k_{\lambda
  z}^{2}\rangle$) represents the average of the operator $-i\partial/\partial
z-K_{\lambda z}^{0}$ [$(-i\partial/\partial z-{K}_{\lambda
  z}^{0})^{2}$] over the electron state of the lowest subband in
$\lambda$ valley. With the infinite-depth assumption, it reads
$\langle k_{\lambda z}\rangle$=0  [$\langle
k_{\lambda z}^{2}\rangle$=$(\pi/a)^{2}$]. The spin-orbit coupling
coefficients utilized in the calculation are
$\gamma=0.011$\ eV$\cdot$nm$^{3}$\cite{aronov,zhou} and
$\beta=0.026$\ eV$\cdot$nm,\cite{fu} respectively. The
interaction Hamiltonian is composed of the intra- ($\lambda^\prime
=\lambda$) and inter-valley ($\lambda^\prime\not=\lambda$)
Coulomb scattering $H_{\lambda\lambda^\prime}^{e-e}$, electron-phonon scattering
$H_{\lambda\lambda^\prime}^{e-p}$ and electron-impurity scattering
$H_{\lambda\lambda^\prime}^{e-i}$.
Expressions of these Hamiltonian
can be found in books\cite{mahan,haug}
and Refs.\ \onlinecite{lei,liu}.

The KSBEs constructed by the nonequilibrium Green's function method
read\cite{wu-rev,wu1}
\begin{equation}
\dot{\rho}_{{\bf k}_{\lambda}}=\left.\dot{\rho}_{{\bf k}_{\lambda}}\right|_{dri}
+\left.\dot{\rho}_{{\bf
    k}_{\lambda}}\right|_{coh}+\sum_{\lambda^{\prime}}\left.
\dot{\rho}_{{\bf k}_{\lambda}}\right|_{scat,\lambda\lambda^{\prime}}\ .
\end{equation}
Here $\rho_{{\bf k}_{\lambda}}$  represent the
density matrices of electrons with relative
 momentum ${\bf k}_{\lambda}$ in valley $\lambda$, whose diagonal terms
$\rho_{{\bf k}_{\lambda},\sigma\sigma}\equiv f_{{\bf
    k}_{\lambda},\sigma}$ ($\sigma=\pm1/2$) represent the electron
distribution functions and the off-diagonal ones $\rho_{{\bf
    k}_{\lambda},\frac{1}{2}-\frac{1}{2}}=\rho_{{\bf  k}_{\lambda},-\frac{1}{2}\frac{1}{2}}^\ast$
 describe the inter-spin-band correlations for the spin coherence.
 $\dot{\rho}_{{\bf k}_{\lambda}}|_{dri}=e{\bf
  E}\cdot\mbox{\boldmath$\nabla$\unboldmath}_{{\bf
    k}_{\lambda}}{\rho}_{{\bf k}_{\lambda}}$
are the driving terms from the external electric field.
$\dot{\rho}_{{\bf k}_\lambda}|_{coh}$ are the coherent terms describing the
coherent spin precessions and $\dot{\rho}_{{\bf k}_{\lambda}}|_{scat,\lambda\lambda^{\prime}}$
stand for the intra-  ($\lambda=\lambda^\prime$) and
inter-valley ($\lambda\ne\lambda^{\prime}$) scattering terms.
 Expressions of these terms are
given in Appendix\ A.

The initial conditions at time $t$=0 are prepared from
spin unpolarized equilibrium states (in the absence of electric field)
at time $t=-t_{0}$:
\begin{eqnarray}
&&\hspace{-0.6cm}f_{{\bf k}_{\Gamma},\sigma}(-t_{0})=\{\exp[(\varepsilon_{{\bf
  k}_{\Gamma}}-\mu)/k_{B}T]+1\}^{-1}\ ,\\
&&\hspace{-0.6cm}f_{{\bf k}_{L_{i}},\sigma}(-t_{0})=
\rho_{{\bf k}_{\Gamma},\frac{1}{2}-\frac{1}{2}}(-t_{0})=
\rho_{{\bf k}_{L_{i}},\frac{1}{2}-\frac{1}{2}}(-t_{0})=0\ ,
\label{eqsum}
\end{eqnarray}
with $\mu$ representing the chemical potential at temperature $T$. We turn on the
electric field at $t=-t_0$ and allow the system to evolve to the steady state
before $t=0$. Under the driving field and the inter-valley
scattering, part of electrons in $\Gamma$ valley are driven
to $L$ valleys. Then we turn on a circularly polarized laser pulse
with width $\delta_{\tau}=0.01$\ ps  at $t=0$ to
excite spin-up electron with Gaussian-like distribution to $\Gamma$
valley: $\delta f_{{\bf k}_{\Gamma},1/2}$=$\alpha\exp[-(\varepsilon_{{\bf
  k}_{\Gamma}}-\varepsilon_{F})^{2}/2\delta_{\varepsilon}^{2}][1-f_{{\bf k}_{\Gamma},1/2}(0)]$.
Here $\alpha=NP_{0}/\{\sum_{{\bf k}_{\Gamma}}
\exp[-(\varepsilon_{{\bf k}_{\Gamma}}-\varepsilon_{F})^{2}/2
\delta_{\varepsilon}^{2}][1-f_{{\bf k}_{\Gamma},1/2}(0)]\}$, $\varepsilon_{F}$ is
the Fermi energy of $\Gamma$ valley and
$\delta_{\varepsilon}$=$\hslash/\delta_{\tau}$. $N$ is the total electron density
after the pulse excitation and
 $P_{0}$ is the spin polarization  excited by the pulse.
In this work, $N=4\times10^{11}$\ cm$^{-2}$ and  $P_{0}=5$\ \%.
\begin{figure}[ht]
    {\includegraphics[width=7.5cm]{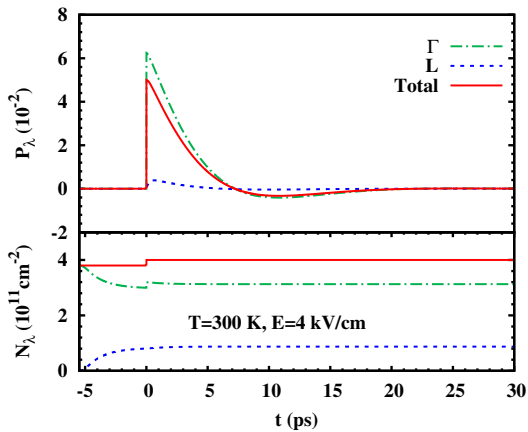}}
    \caption{(Color online) A typical time evolution of the initial
      steady-state preparation (density $N_\lambda$ and
      spin polarization $P_\lambda$)
      under the electric field $E=4$ kV/cm and a spin-polarization generation by a
      laser pulse at $t=0$. $N_i=B=0$. Chain curves: $\Gamma$
      valley; Dashed curves: $L$ valleys; Solid curves: the total.}
  \label{fig1}
\end{figure}

\section{Numerical results}
\begin{figure}[ht]
    {\includegraphics[width=9cm]{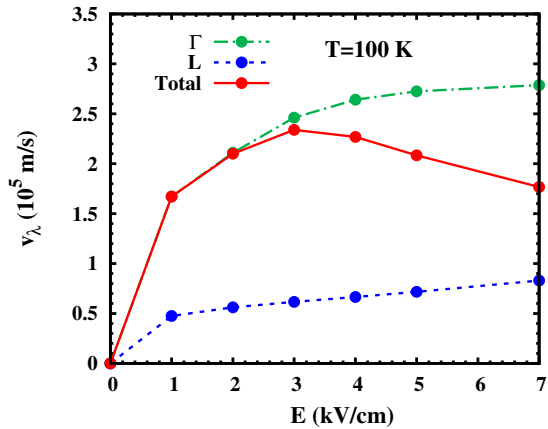}}
    \caption{(Color online) Steady-state drift velocity $v_{\lambda}$
      against electric field $E$ at $T$=100\ K. $N_i=B=0$. Chain curve:
      $\Gamma$ valley; Dashed curve: $L$
      valleys; Solid curve: the total.}
  \label{fig2}
\end{figure}

We numerically solve the KSBEs following the scheme mainly laid out in Ref.\ \onlinecite{weng1},
with extension to the inter-valley scattering given in Appendix\ B.
In Fig.\ \ref{fig1} we show a typical time evolution of the initial steady-state preparation
under the influence of the electric field and a spin-polarization generation by a
laser pulse. The electric field is
applied from time $t=-t_{0}=-5.5$\ ps, and the laser pulse is applied
at $t=0$\ ps.
$P_{\lambda}(t)\equiv2\sum_{{\bf k}_{\lambda}\sigma}[\sigma f_{{\bf
      k}_{\lambda},\sigma}(t)]/\sum_{{\bf k}_{\lambda}\sigma}f_{{\bf
      k}_{\lambda},\sigma}(t)$ and $N_{\lambda}(t)\equiv\sum_{{\bf k}_{\lambda}\sigma}f_{{\bf
      k}_{\lambda},\sigma}(t)$ are
the spin polarization and the electron density at $\lambda$ valley, separately.
 The steady-state drift velocity of $\lambda$
  valley ${\bf v}_{\lambda}$ can be obtained from the steady value
of ${\bf v}_{\lambda}(t)\equiv\sum_{{\bf k}_{\lambda}\sigma}[f_{{\bf
      k}_{\lambda},\sigma}(t)\hslash{{\bf k}_{\lambda}}/m_{\lambda}]/\sum_{{\bf k}_{\lambda}\sigma}f_{{\bf
      k}_{\lambda},\sigma}(t)$. The steady-state drift velocities
  $v_{\lambda}$ of each valley as well as the total drift velocity
against the electric field $E$ are plotted in
Fig.\ \ref{fig2}. From the figure the negative differential
electric conductance of the total drift velocity is obtained.

\begin{figure}[ht]
    {\includegraphics[width=9cm]{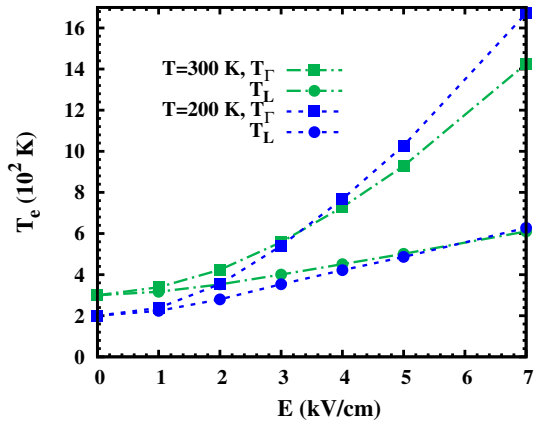}}
    \caption{(Color online) Hot-electron temperature $T_{\Gamma}$
      (curves with squares) and $T_{L}$ (curves with circles) against
      electric field $E$ at lattice temperature $T$=300\ K (chain
      curves) and 200\ K (dashed curves). $N_i=B=0$.}
  \label{fig3}
\end{figure}

The hot-electron temperature $T_{e}$ can be
obtained by fitting the Boltzmann tail of the calculated steady-state electron
distribution of each valley.\cite{weng1} Our results show that each
valley has its own hot-electron temperature(s).
Moreover, for $\Gamma$ valley, there are two temperatures.
One is at the energy regime which overlaps with that of $L$ valleys. In this regime,
electrons at $\Gamma$ and $L$ valleys share the {\em same}
hot-electron temperature (labeled as $T_{L}$) due to the strong inter-valley
scattering. The other is $T_\Gamma$ for electrons in the lower energy
regime of $\Gamma$
valley. That is the typical hot-electron temperature of $\Gamma$ valley and is
higher than $T_L$. $T_{\Gamma}$ and $T_{L}$ as functions of electric field
at different lattice temperatures are plotted in Fig.\ \ref{fig3}.
It shows that electrons with smaller effective mass (such as those of $\Gamma$
valley) and lower lattice temperature (thus relatively weak scattering)
are easier to be accelerated and heated.

The spin relaxation time $\tau$ can be obtained from the temporal evolution of spin
polarization $P_\lambda$.  Due to the lowest conduction subbands used
in each valley and  the ``run-away'' effect,\cite{dmitriev}
our research is limited to electric field up to 7\ kV/cm.

\begin{figure}[ht]
    {\includegraphics[width=9cm]{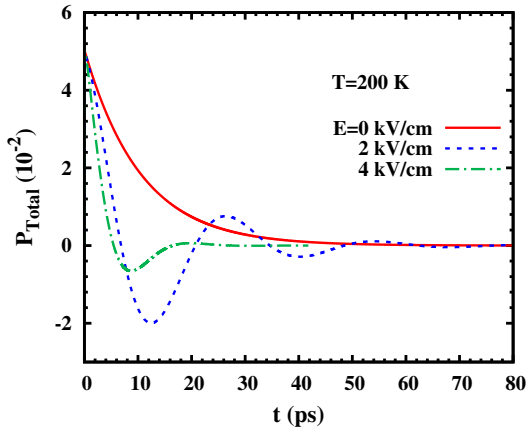}}
    \caption{(Color online) Temporal evolution of spin polarization
      $P_{Total}$ when $E=0$ (solid curve), 2
      (dashed curve), and 4\ kV/cm (chain curve). The corresponding spin relaxation times
are fitted to be 10.5, 14.4, and 4.7\ ps, respectively. $N_i=B=0$.}
  \label{fig4}
\end{figure}

\subsection{Temporal evolution of spin polarization}
We first show the temporal evolutions of total spin polarization $P_{Total}(t)\equiv
\sum_\lambda P_\lambda(t)$ under different
electric fields at $T=200$\ K with $B=N_{i}=0$
in Fig.\ \ref{fig4}. The corresponding
spin relaxation times obtained from the calculations are also given in the
figure caption. It is seen from the
figure that in the presence of electric field, electron spins
precess with the precession period showing $E$-dependence. This is because
the center-of-mass momentum induced by the electric field leads to an effective
magnetic field via the Dresselhaus spin-orbit coupling Eqs.\
(\ref{hg}) and (\ref{eqso}).\cite{weng1} Another interesting
phenomenon is that the spin relaxation first decreases and then increases
with electric field. It is known that the
electric field can cause two
effects: hot-electron
effect and center-of-mass drift effect.\cite{weng1}
The hot-electron effect increases the inhomogeneous broadening as
well as the scattering. The joint effects lead to
enhanced (reduced) spin R\&D in weak
(strong) scattering limit,\cite{lv,zhou}
when the linear Dresselhaus term of the $\Gamma$ valley is important.\cite{weng2}
The second effect forces spins to
precess around the induced magnetic
field mentioned previously and hence reduces the inhomogeneous broadening
and the spin R\&D. However, this effect is quite marginal compared
to the first one.\cite{weng1}
It has been demonstrated before that electrons in $\Gamma$ valley is
in the strong scattering limit.\cite{zhou,weng1,lv}
 When the electric field
is low, electrons are mainly distributed in the $\Gamma$ valley. Thus the
influence of hot-electron effect in the strong
scattering limit causes $\tau$ to increase with $E$. However,
when the electric field is high enough, apart from the fact that the
increase of inhomogeneous broadening from the cubic
term of the Dresselhaus term in $\Gamma$ valley
becomes important,\cite{weng2}   $L$ valleys also start to play an
important role. The amount of electrons sitting in $L$ valleys
grows with the electric field and the hot-electron temperature increases
with the field as well. Thus the inter-valley
scattering becomes more important with the
increase of electric field.
The strong spin-orbit coefficient at $L$ valleys
in conjunction with the inter-valley scattering
lead to the decrease of the spin relaxation time
with the field.

\begin{figure}[htb]
    {\includegraphics[width=8cm]{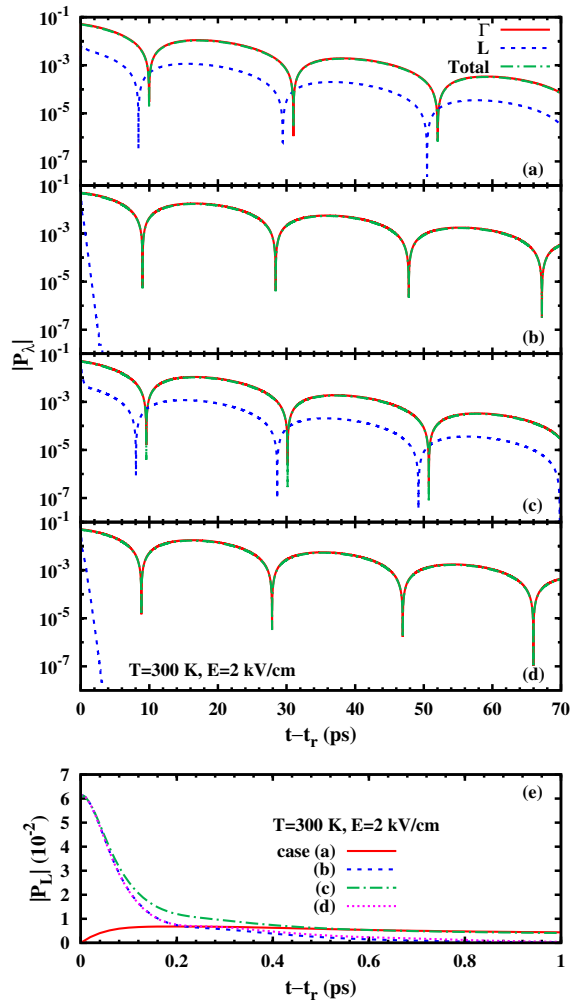}}
    \caption{(Color online) Spin polarization
      $|P_{\lambda}|$ {\em vs.} $t-t_r$ under electric field $E=2$\ kV/cm. Solid curve:
      $\Gamma$ valley; dashed
      curve: $L$ valley; and chain curve: the total. (a):
      with both $H_{\Gamma-L}^{e-p}$ and $H_{\Gamma-L}^{e-e}$;
      (b): without $H_{\Gamma-L}^{e-p}$; (c): without
      $H_{\Gamma-L}^{e-e}$; and (d): without $H_{\Gamma-
        L}^{e-e}$ and $H_{\Gamma-L}^{e-p}$. $T=300$\ K and
      $N_i=B=0$. (e): The initial time evolution of spin polarization
      of $L$ valleys.}
  \label{fig5}
\end{figure}

\subsection{Effect of inter-valley scattering on spin relaxation}
To elucidate the effect of inter-valley scattering on spin
relaxation, we compare the spin relaxation under four different
conditions: (a) with both $\Gamma$-$L$ inter-valley Coulomb and
electron-phonon scattering, (b)
with $\Gamma$-$L$
inter-valley Coulomb scattering $H_{\Gamma-L}^{e-e}$ only,
(c) with $\Gamma$-$L$
inter-valley electron-phonon scattering $H_{\Gamma-L}^{e-p}$ only
 and (d) without any $\Gamma$-$L$ inter-valley
scattering. (It is stressed that besides  $H_{\Gamma-L}^{e-e}$ and
 $H_{\Gamma-L}^{e-p}$, all the other scattering terms are always present.)
In order to get a clear insight into the inter-valley scattering to the
spin relaxation, we need to prepare enough electron density and spin polarization
in the $L$-valleys. To do so, we first switch off the
coherent terms at time $t=0$ (or
even at $t=-t_0$, both yield the same result where there is
no spin polarization) and
allow the system to evolve to a spin polarized steady state
(typically, the time $t_r$ needed to reach a steady spin polarization  is about
6\ ps after the circularly polarized laser pulse is applied when $T=300$\ K,
$E=2$\ kV/cm, and $B=N_i=0$). It is
noted that all inter-valley scattering should be included
at this stage  in order to allow
$\Gamma$ electrons enter into $L$ valleys, driven by the electric field.
Then we switch on the coherent terms to allow spin precession and
switch off the corresponding $\Gamma$-$L$ inter-valley scattering at
time $t_r$. One should keep in mind  that in our full calculation [such as under
condition (a)], the coherent terms and all the scattering terms are
present at all the time.

In Fig.\ \ref{fig5} we plot time evolution of $|P_{\lambda}|$ against $t-t_r$
starting from the moment when the spin relaxation begins.
$t_r$ is taken as 0 for condition (a). In the calculation, $T=300$\ K,
$E=2$\ kV/cm and $B=N_i=0$.
From Fig.\ \ref{fig5}(a), one can see that
in the real situation, $\Gamma$ and $L$ valleys share almost the identical damping
rate and precession period in spin precession, although the initial
spin polarization of $L$ valleys is very small. However, when
$H_{\Gamma-L}^{e-p}$ is absent, {\em huge} differences between spin
relaxations in $\Gamma$ and $L$ valleys appear: Despite whether $H_{\Gamma-
  L}^{e-e}$ is present [Fig.\ \ref{fig5}(b)] or absent [Fig.\ \ref{fig5}(d)],
the spin relaxation in $L$ valleys is
much faster than that in $\Gamma$ valley, regardless of much larger initial spin polarization
now gained by the  $L$ valleys  due to our controlling
trick presented above [it is shown in Fig.\ \ref{fig5}(e) that $P_L(0)$ can even be
higher than 5\ \%]. Furthermore, there is no spin precession
in $L$ valleys anymore.  It is further noted that the $\Gamma$-$L$ inter-valley
electron-phonon scattering is the dominant inter-valley scattering. This can be seen from
Fig.\ \ref{fig5}(c) that in the presence of $H_{\Gamma-L}^{e-p}$  but in the
absence of $H_{\Gamma-L}^{e-e}$, the
spin polarization of $L$ valleys shows the same damping rate and  precession period.
This is because when the electric field is relatively low (here $E=2$\ kV/cm), only a small
number of electrons distribute in $L$ valleys [see the inset of
Fig.\ \ref{fig8}(a)]. Therefore the  inter-valley Coulomb
scattering is very weak.

\begin{figure}[htb]
    {\includegraphics[width=9cm]{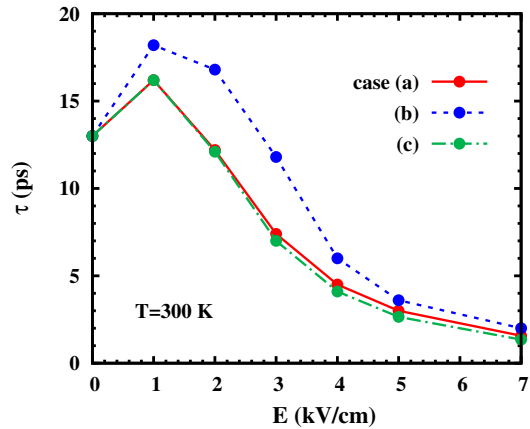}}
    \caption{(Color online) Spin relaxation time $\tau$ {\em vs.}
      electric field $E$. Solid curve:
      with both $H_{\Gamma-L}^{e-p}$ and $H_{\Gamma-L}^{e-e}$ [case (a)];
Dashed curve: without $H_{\Gamma-L}^{e-p}$ [case
      (b)]; Chain curve: without $H_{\Gamma-L}^{e-e}$ [case (c)].
      $T=300$\ K and $N_i=B=0$.}
  \label{fig6}
\end{figure}

The above features can be well understood with respect to the
different spin-orbit couplings in  different valleys, in conjunction
with the strong inter-valley
scattering. The spin-orbit coupling strength of $L$ valleys
is about 16 times as large as that of
$\Gamma$ valley.\cite{comp}  This strong
inhomogeneous broadening leads to a fast
spin decay in $L$ valleys if they are isolated from $\Gamma$ valley [see the
dashed curve in Fig.\ \ref{fig5}(d), which has a corresponding spin
relaxation time in the order of 0.1\ ps]. Moreover, the fast spin relaxation of $L$
valleys also manifests itself in the quick initial spin decay after turning on the
coherent terms in Fig.\ \ref{fig5}(b-d),
as shown in Fig.\ \ref{fig5}(e). It is therefore understood that due
to the efficient inter-valley exchange of electrons
caused by the inter-valley electron-phonon
scattering as well as the large spin-orbit coupling in $L$ valleys,
 $L$ valleys play the role of a ``drain'' of spin polarization
and enhance spin relaxation of the total system effectively [comparing
the chain curves in Fig.\ \ref{fig5}(a) and Fig.\ \ref{fig5}(b)].  Meanwhile,
 the strong inter-valley electron-phonon scattering causes the spin evolutions
of different valleys to be almost identical.

To have a further insight into the effect of inter-valley scattering on
spin relaxation under different electric fields, the spin relaxation time
$\tau$ against $E$ is shown in Fig.\ \ref{fig6} under
conditions (a), (b) and (c) defined above. In the figure, the
enhancement of spin relaxation caused by
$L$ valleys via inter-valley electron-phonon scattering is obvious
by comparing the solid and dashed curves. The effect of inter-valley Coulomb
scattering becomes visible when $E$ is high ($>2$\ kV/cm), leading to
a marginal decrease of the spin relaxation. That is because in the strong
scattering limit, adding new scattering tends  to reduce the spin
relaxation by its counter effect on inhomogeneous broadening.\cite{lv}

\begin{figure}[ht]
    {\includegraphics[width=9cm]{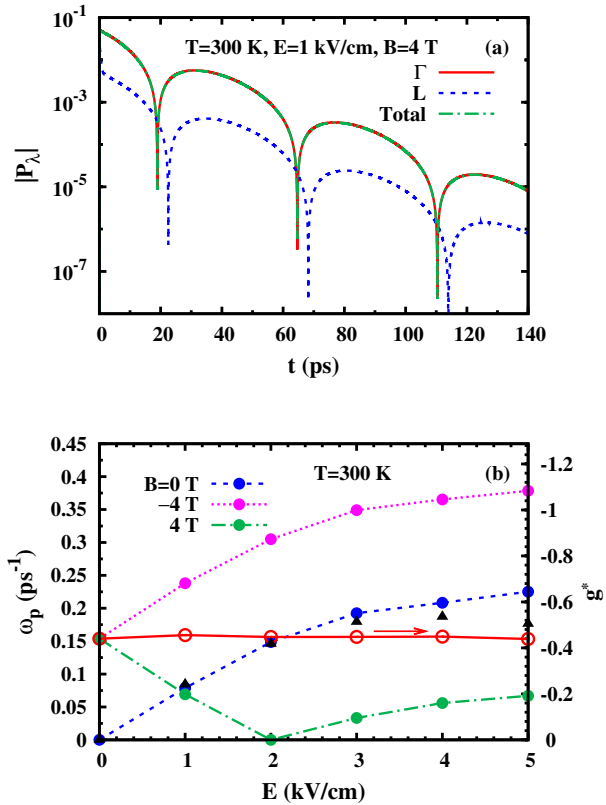}}
    \caption{(Color online) (a): Temporal evolution of $|P_{\lambda}|$
      with electric field $E=1$\ kV/cm and magnetic field $B=4$\
      T. (b): Spin precession frequency
      $\omega_{p}$ against electric field $E$ when static magnetic field
      $B=0$\ T (dashed curve), $-4$\ T (dotted curve) and $4$\
      T (chain curve). Triangular dots: spin precession
      frequency under electric field predicted by Eq.\ \ref{analyze} when $B=0$\ T.
      Solid curve: calculated
      effective $g$-factor $g^\ast$ under different electric fields
      (note the scale of this curve is on the right hand side of the frame).}
  \label{fig7}
\end{figure}

\subsection{Spin precession}

Now we turn to investigate the property of spin precession. It is
known that in the presence of electric field, spins can precess under
the electric-field-induced effective magnetic field.\cite{weng1}
When a static magnetic field is further applied,
spins then precess around the combined effective magnetic field. We
define the combined effective magnetic field term as
${\bf \Omega}({\bf B})={\bf \Omega}_0+\frac{1}{2}g^{\ast}\mu_{B}{\bf B}$. Here
${\bf \Omega}_0$
is the induced magnetic field term, which is dependent
on the electric field. $g^{\ast}$ is  the effective $g$-factor. The
corresponding spin precession frequency around ${\bf \Omega}({\bf B})$
is $\omega_{p}\equiv2|{\bf \Omega}({\bf B})|/\hslash$. For each valley,\cite{weng1}
\begin{equation}
{\bf \Omega}_0^\lambda=\frac{\int d{\bf k}_{\lambda}(f_{{\bf
  k}_{\lambda},\frac{1}{2}}-f_{{\bf
  k}_{\lambda},-\frac{1}{2}}){\bf h}_{\lambda}({\bf k}_{\lambda})}{\int
d{\bf k}_{\lambda}(f_{{\bf k}_{\lambda},\frac{1}{2}}-f_{{\bf
    k}_{\lambda},-\frac{1}{2}})}\ .
\label{omega0}
\end{equation}
From Eqs.\ (\ref{hg}) and (\ref{eqso}), one finds immediately
that for $\Gamma$ valley, ${\bf \Omega}_0^{\Gamma}$ is along the
$x$-axis when the electric field is along the $x$-axis (note that $\langle
k_{\Gamma z}^2\rangle$ is larger than the mean
value of $k_{\Gamma y}^2$ due to the strong confinement along the $z$-axis),
while for $L$ valleys, \{${\bf \Omega}_0^{L_i}$\} share the same
magnitude but direct along (0, $\pm$1, $\pm$1) directions.
Therefore there is no net induced magnetic field
under electric field in $L$-valleys, i.e., $\sum_i{\bf \Omega}_0^{L_i}=0$ and
${\bf \Omega}_0$ of the whole system comes
from ${\bf \Omega}_{0}^{\Gamma}$. A direct inferrer  is that
no spin precession will appear for electrons in $L$ valleys under electric field only.
However, as shown in Fig.\ \ref{fig5}(a), due to the strong $\Gamma$-$L$ inter-valley scattering, spin
precession in $L$ valleys under electric field only is observed and
shares the {\em same} precession period as the $\Gamma$ valley. One may also
expect that when a static magnetic field is applied, electron spins at $L$ valleys may have a
distinct spin precession frequency as
the $g$-factor of $L$ valley is three times as large as that of $\Gamma$ valley.
 Nevertheless, again this is not the case. Due to both the
strong inhomogeneous broadening (i.e., fast spin relaxation in $L$
valleys) and the intensive exchange of electrons between $\Gamma$ and $L$ valleys
due to the strong inter-valley scattering, $L$ valleys still present the same
precession period as that of $\Gamma$ valley,
 as shown in Fig.\ \ref{fig7}(a). This is because electron spins in $L$ valleys
can hardly show any spin precession characterized by $g_L\mu_B B/2$
without being depolarized from the strong inhomogeneous broadening-induced interference decay as well
as being scattered back to $\Gamma$ valley due to the strong
inter-valley scattering. (In fact, the spin precession time characterized
 by $g_L\mu_B B/2$ is 13.2\ ps, the spin precession time
under the momentum-dependent effective magnetic field from
  spin-orbit coupling in $L$ valleys is around 0.2\ ps and the $L$-$\Gamma$
inter-valley electron-phonon scattering time calculated using
Fermi's golden rule is about 0.4\ ps when
  $T=300$\ K.)
Therefore, the spin polarization in $L$ valleys is just ``pumped '' from the $\Gamma$ valley
due to the strong inter-valley scattering. Consequently, the
spin precession of the total system is determined by
$\Gamma$ valley. In the following, we prove this by
exploring the effective $g$-factor of the total system.

The frequency of spin precession $\omega_{p}$ as function of electric
filed $E$ is plotted in Fig.\ \ref{fig7}(b), in both the presence and absence of
the applied magnetic field in Voigt configuration. It is shown that when
the magnetic field is
absent or is applied along $-x$-axis direction, $\omega_{p}$ increases monotonically with the
electric field. Nevertheless, when the magnetic field is along
$x$-axis, $\omega_{p}$ first decreases
 to zero and then increases again with $E$. The monotonic increase in the absence of the magnetic
field is due to the increase of the electric-field-induced magnetic field. In fact, when the spin
polarization is small, in low electric field regime the induced
 magnetic field term ${\bf \Omega}_0$ can be roughly estimated by Eq.\
(\ref{omega0}) of $\Gamma$ valley, i.e.,
\begin{equation}
{\bf \Omega}_0=\frac{\gamma
  m_\Gamma^2{\bf v}_\Gamma}{\hslash^3}\Big[\frac{\varepsilon_F}
{2(1-e^{-\varepsilon_F/k_B T_\Gamma})}-\frac{\hslash^2\pi^2}{2m_\Gamma a^2}\Big]\ .
\label{analyze}
\end{equation}
The spin precession frequency calculated using Eq.\ (\ref{analyze})
in the absence of magnetic field is plotted as triangular dots in Fig.\ \ref{fig7}(b). It is
seen that they coincide pretty well with the full
calculation at low electric field regime.
As the induced magnetic field is along $-x$-axis, consequently an applied
  static magnetic field along $-x$-axis increases the spin precession
frequency universally by a constant number (comparing the dashed and
dotted curves in the figure). However, when the applied
magnetic field is along $x$-axis which is in
opposite direction to the induced magnetic field, the induced
magnetic field does not overcome the static magnetic field until the
electric field is larger than 2\ kV/cm. By utilizing the relation
$-\frac{1}{2}g^{\ast}\mu_{B}B|_{B=4}=\Omega(0)-\Omega(4)$ (or $\frac{1}{2}g^{\ast}\mu_{B}B|_{B=-4}=\Omega(-4)-\Omega(0)$),
one can get $g^{\ast}$  under different
electric fields. Simple calculations show that
$g^\ast$ changes little with electric field and is around the value
when the electric field is abscent, i.e., $g^\ast\approx
g_\Gamma=-0.44$ [see the solid curve in Fig. \ref{fig7} (b)],
 suggesting that the spin precession
properties of the total system mainly come from electrons of $\Gamma$ valley.

\subsection{Effect of temperature, magnetic field, and  impurity on
  spin relaxation}

\begin{figure}[ht]
    {\includegraphics[width=8cm]{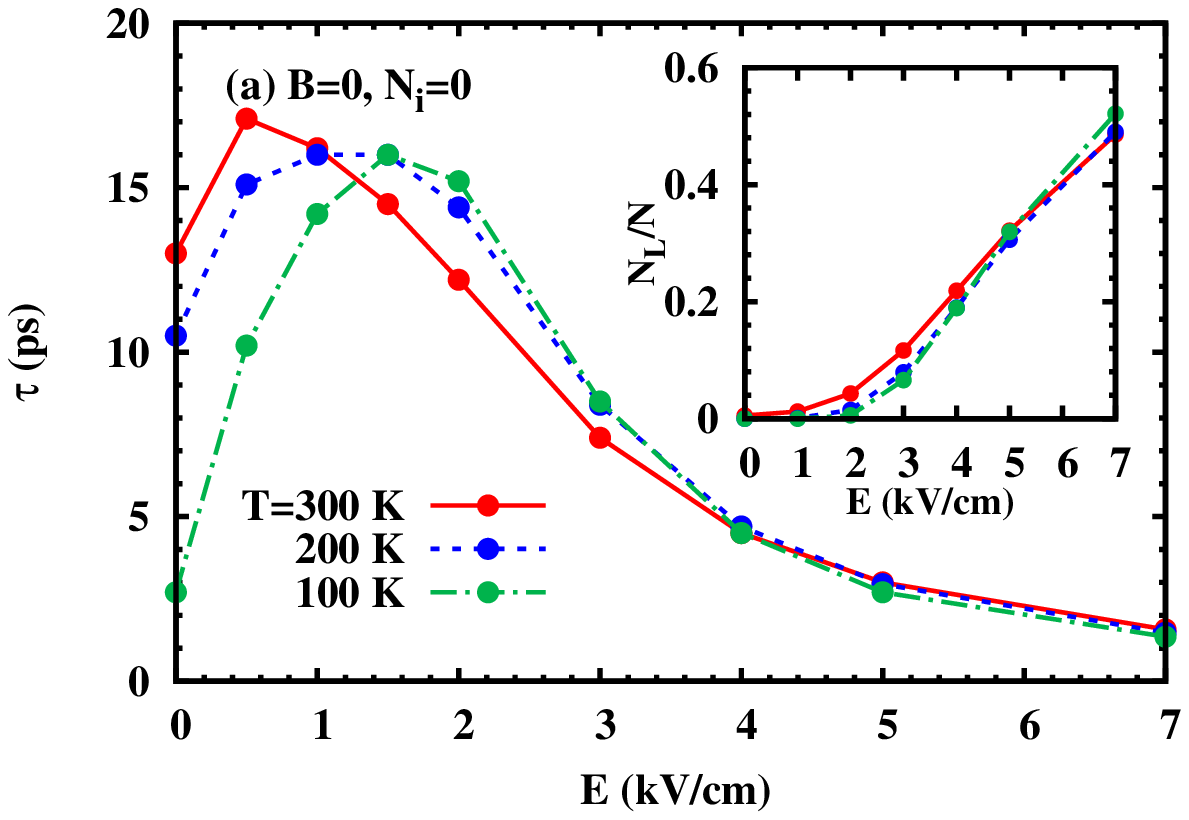}}
    {\includegraphics[width=8cm]{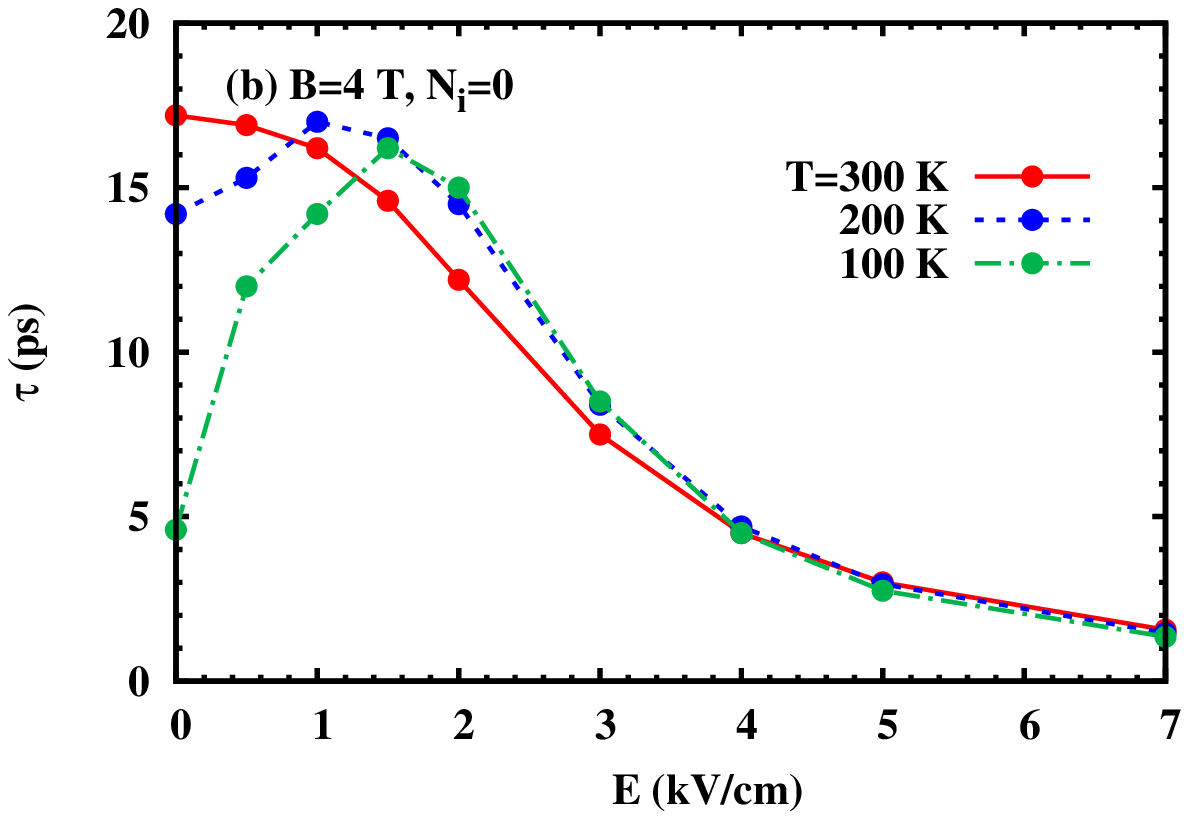}}
    {\includegraphics[width=8cm]{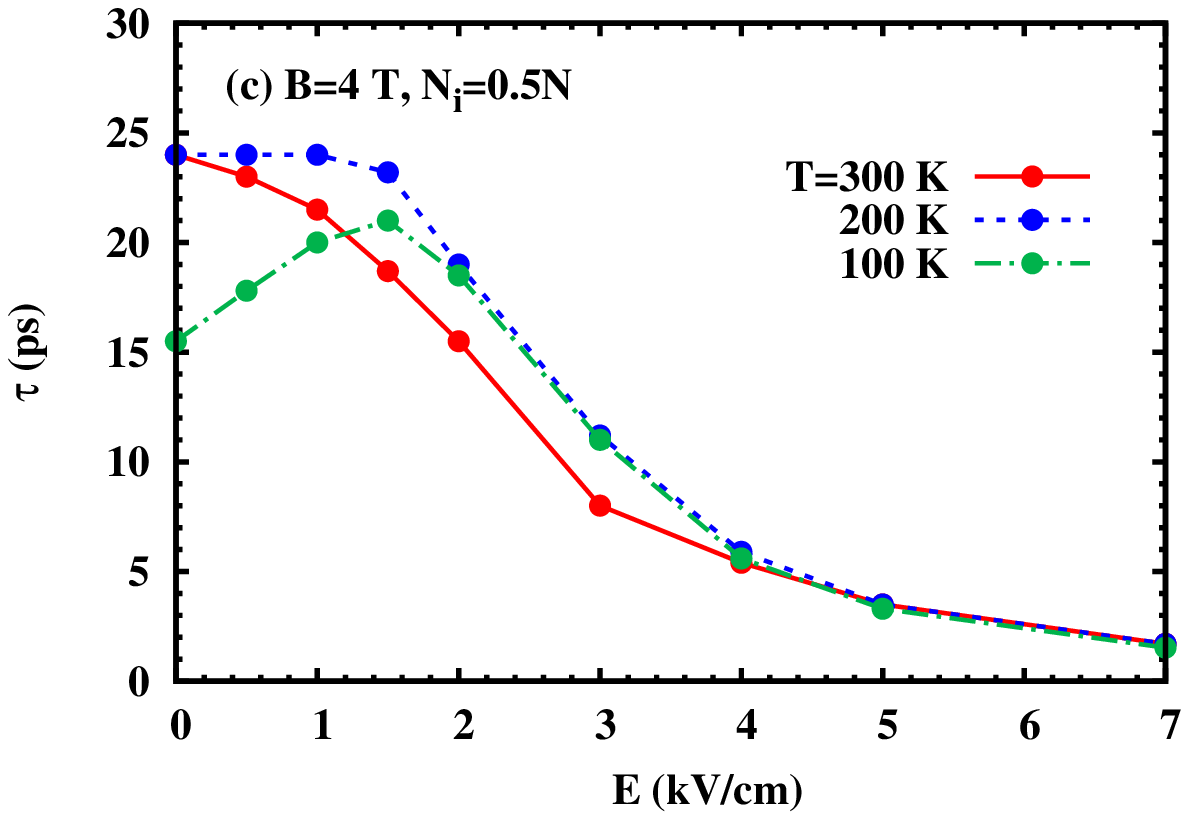}}
    \caption{(Color online) Spin relaxation time $\tau$ against electric filed $E$ when
temperature $T$=300 (solid curves), 200  (dashed curves) and
100\ K (chain curves) under different conditions:
(a) $B=N_i=0$; (b) $B=4$\ T and $N_i$=0; and (c) $B=4$\
      T and $N_i=0.5N$. Inset of (a): fraction of electrons in $L$
      valleys against electric field.}
    \label{fig8}
\end{figure}

In this subsection  we discuss the effect of temperature, magnetic field, and
impurity on spin relaxation. Spin relaxation time $\tau$ against
electric filed $E$ at
different temperatures $T=300$, 200 and 100\ K with and
 without magnetic field/impurities are
shown in Fig.\ \ref{fig8}.

From the figure one can see that in each case, the peak in $\tau$-$E$ curve
(first discussed in Sec.\ IIIB) is
more pronounced and shows up at a relatively higher
electric field when $T$ is decreased. This is
due to two facts: First, the hot-electron effect mentioned
previously in Sec.\ IIIA is more
important when $T$ is low.\cite{weng1} Second, when temperature is low, the $\Gamma$-$L$ inter-valley exchange
of electrons in low electric
field regime is marginal due to the low
electron density in
$L$ valleys [see the inset of Fig.\ \ref{fig8}(a)]. Thus the enhancement
of spin relaxation by $L$ valleys does not come into effect markedly
until the electric field is high enough.

It is also
shown from the figure that in the low electric field
regime ($E\leqslant 2$\ kV/cm),
the spin relaxation can be reduced by
increasing temperature and/or magnetic field and/or impurity density.
 The reduction of the spin relaxation due to higher temperature and
impurity density
corresponds to the hot-electron
effect in strong scattering limit---the increase
of temperature and/or adding more impurities strengthen
the scattering, thus reduce the spin
R\&D. The main effect of static magnetic field in the low electric
field regime is to reduce the spin R\&D by forcing spins to undergo a
Larmor precession around it and thus reducing inhomogeneous broadening.\cite{weng3}

Finally one notices from the figure that when $E$ is high enough, the
effects of lattice temperature, impurity density, and magnetic field are
marginal. In this regime, the inter-valley electron-phonon scattering is
considerably strengthened due to the high hot-electron temperatures, and thus dominates and
becomes insensitive to
the background temperature. Thus the changes in lattice temperature and other
kinds of scattering cause little
discrepancy in spin relaxation. Meanwhile, the inhomogeneous broadening is highly
enhanced as well in this high electric field regime. Therefore the finite magnetic
field has little restriction on inhomogeneous broadening and affects spin
relaxation marginally. Consequently, the spin relaxation under high
electric field tends to be identical and insensitive to the lattice
temperature, magnetic field and impurity density.

\section{Conclusion}

In conclusion, we investigate the multi-valley spin relaxation in
$n$-type GaAs QWs in the presence of in-plane
electric field under high temperature ($\geqslant100$\ K),
by applying the KSBE approach. Due to the small well width in this
investigation, electrons at $\Gamma$ and $L$ valleys determine the
properties of spin dynamics at high electric field.
We discuss the effect of inter-valley
scattering on spin relaxation, the spin precession properties, and the
spin relaxation under different conditions, such as lattice
temperatures, magnetic fields and impurity densities.

First, the negative differential electric conductance, which is
essential for the charge Gunn effect, is reproduced.
Meanwhile, the hot-electron temperature is also
investigated, showing that $\Gamma$ valley has {\em two} distinct
temperatures. One is for electrons in the energy regime which overlaps
with that of $L$ valleys, and shows the {\em same} hot-electron
temperature of those in $L$ valleys ($T_L$). The other  is for electrons in the
lower energy regime of $\Gamma$ valley ($T_\Gamma$). It is shown that
$T_\Gamma>T_L$. The identical electron
temperature in the overlapped energy regime of $\Gamma$ and $L$
valleys is due to the strong
inter-valley scattering whereas $T_\Gamma>T_L$ results from the smaller
effective mass in $\Gamma$ valley.

The spin relaxation time $\tau$ first increases and then decreases with the electric field,
especially at low temperatures. The first increase of $\tau$ in low
electric field regime is due to the hot-electron effect in the strong scattering
limit. Whereas the following
decrease of $\tau$ in the high electric field regime is mainly due to
the fast spin relaxation in $L$ valleys (originating from the much
stronger spin-orbit coupling at $L$ valleys) and strong $\Gamma$-$L$
inter-valley scattering. In fact, $L$ valleys are shown to play the
role of a ``drain'' of the total spin polarization. It is also shown that the
$\Gamma$-$L$ inter-valley electron-phonon scattering is responsible for the
inter-valley scattering.
The contribution of $\Gamma$-$L$ inter-valley electron-electron Coulomb
scattering is marginal as it does not cause any
electron exchange between different valleys. Moreover, due to the
strong $\Gamma$-$L$ inter-valley scattering as well as the
strong inhomogeneous broadening in $L$ valleys,
electrons in $L$ valleys possess almost the same spin relaxation rate and spin precession period
as those in $\Gamma$ valley. It is further shown
that this feature still holds when a static magnetic field is
applied. Actually, in our investigation ($B\leq 4$\ T), the spin precession frequency is
determined by electrons at $\Gamma$ valley, despite the fact that $g_L$ is much larger in magnitude
than $g_\Gamma$. This is because
the net electric-field-induced magnetic field from $L$ valleys  is zero,
 and the effect of static magnetic field in $L$ valleys
can not surpass the much stronger inhomogeneous broadening as well as
frequent exchange of electrons from $\Gamma$ valley.

We also investigate the effects of temperature, magnetic field and
impurity on spin relaxation. When the electric field is not too high,
the spin relaxation can be reduced by increasing temperature,
magnetic field, and impurity density.  However,
when the electric field is high enough, spin relaxation time is
dominated by the $\Gamma$-$L$ inter-valley electron-phonon scattering and
becomes insensitive to temperature,
impurity density and magnetic field.

Finally we remark on the possibility of the ``spin Gunn effect'' in
GaAs QWs. First, as pointed out in our previous work,\cite{wu-tran}
 the feasibility of drift-diffusion
model used in Ref.\ \onlinecite{flatte} is valid only in system without
Dresselhaus/Rashba spin-orbit coupling. For Zinc-blende semiconductors,
even without magnetic field, electron spins can
flip due to the Dresselhaus/Rashba spin-orbit coupling, let alone an effective magnetic field induced by electric
field via the same spin-orbit coupling. Therefore, although with the spin-dependent mobilities,
electrons with opposite spins can experience frequent spin-flip.
Second, our calculations show that in
high electric field regime, spin relaxation time decreases effectively with
electric field. This is due to the hot-electron effect,
together with the fast spin relaxation in the upper
valleys.  Therefore, the condition for ``spin Gunn effect'' proposed in
Ref.\ \onlinecite{flatte} can hardly be realized due to the short spin
relaxation time in GaAs QWs.

\begin{acknowledgments}
This work was supported by the Natural Science Foundation of China
under Grant Nos.\ 10574120 and 10725417, the National Basic Research Program of
China under Grant No.\ 2006CB922005 and the Knowledge Innovation
Project of Chinese Academy of Sciences. One of the authors
(P.Z.) would like to thank M. Q. Weng and J. H. Jiang for valuable
discussions.
\end{acknowledgments}

\begin{appendix}
\section{The coherent and scattering terms of the KSBEs}

The  coherent terms are
\begin{equation}
  \dot{\rho}_{{\bf k}_{\lambda}}|_{coh}=-i\big[{\bf\Omega}_{\lambda}({\bf
    k}_{\lambda})\cdot{\bgreek \sigma}-\sum_{\bf
    q}V_{\lambda\lambda,{\bf q}}\rho_{{\bf k}_{\lambda}-{\bf
      q}},\rho_{{\bf k}_{\lambda}}\big]\ ,
\end{equation}
with ${\bf \Omega}_{\lambda}({\bf k}_{\lambda})\equiv{\bf h}_{\lambda}({\bf
  k}_{\lambda})+\frac{1}{2}g_{\lambda}\mu_{B}{\bf B}$ and $V_{\lambda\lambda,{\bf
q}}$ representing the intra-valley Coulomb scattering matrix
element. $[A, B]$ stands for the commutator of $A$ and $B$.
 The scattering terms are
\begin{widetext}
\begin{equation}
\dot{\rho}_{{\bf
    k}_{\lambda}}|_{scat,\lambda\lambda^{\prime}}=-\{S^{{\bf k}_{\lambda}}_{\lambda\lambda^{\prime}}(>,<)
-S^{{\bf k}_{\lambda}}_{\lambda\lambda^{\prime}}(<,>)
+S^{{\bf k}_{\lambda}}_{\lambda\lambda^{\prime}}(>,<)^{\dagger}-S^{{\bf k}_{\lambda}}_{\lambda\lambda^{\prime}}(<,>)^{\dagger}\}\ ,
\end{equation}
where
\begin{eqnarray}\nonumber
S^{{\bf k}_{\lambda}}_{\lambda\lambda^{\prime}}(>,<)&=&\pi
N_{i}\delta_{\lambda\lambda^{\prime}}\sum_{\bf q}U^{2}_{\lambda\lambda,\bf q}\rho_{{\bf
    k}_{\lambda}-{\bf q}}^{>}\rho_{{\bf
    k}_{\lambda}}^{<}\delta(\varepsilon_{{\bf
    k}_{\lambda}}-\varepsilon_{{\bf k}_{\lambda}-{\bf q}})\\\nonumber
&&\mbox{}+\pi\sum_{{\bf q},{\bf
    k}^{\prime}_{\lambda^{\prime}}}V_{\lambda\lambda^{\prime},\bf
  q}^{2}\rho_{{\bf k}_{\lambda}-{\bf q}}^{>}\rho_{{\bf
    k}_{\lambda}}^{<}\mbox{Tr}[\rho_{{\bf
    k}^{\prime}_{\lambda^{\prime}}}^{>}\rho_{{\bf
    k}^{\prime}_{\lambda^{\prime}}-{\bf q}}^{<}]
\delta(\varepsilon_{{\bf k}_{\lambda}-{\bf
    q}}-\varepsilon_{{\bf k}_{\lambda}}+\varepsilon_{{\bf
    k}^{\prime}_{\lambda^{\prime}}}-\varepsilon_{{\bf
    k}^{\prime}_{\lambda^{\prime}}-{\bf q}})\\
&&\mbox{}+\pi\sum_{{\bf
    k}_{\lambda^{\prime}}^{\prime}}M^{2}_{\lambda\lambda^{\prime},\bf q}\rho_{{\bf
    k}^{\prime}_{\lambda^{\prime}}}^{>}\rho_{{\bf
    k}_{\lambda}}^{<} [N_{\lambda\lambda^{\prime}}^{<}
\delta(\varepsilon_{{\bf k}_{\lambda}}-\varepsilon_{{\bf
    k}_{\lambda^{\prime}}^{\prime}}-\Omega_{\lambda\lambda^{\prime}})
+N_{\lambda\lambda^{\prime}}^{>}\delta(\varepsilon_{{\bf k}_{\lambda}}
-\varepsilon_{{\bf  k}_{\lambda^{\prime}}^{\prime}}+\Omega_{\lambda\lambda^{\prime}})]\ .
\end{eqnarray}
\end{widetext}
Here $\rho_{\bf k}^{<}=\rho_{\bf k}$ and $\rho_{\bf
  k}^{>}=1-\rho_{\bf
  k}$. $N_{\lambda\lambda^{\prime}}=N_{\lambda^{\prime}\lambda}=[\exp(\hslash\Omega_{\lambda\lambda^{\prime}}/k_{B}T)-1]^{-1}$
is the Boson distribution of phonons with frequency
$\Omega_{\lambda\lambda^\prime}$. $N^<_{\lambda\lambda^\prime}=N_{\lambda\lambda^\prime}$ and
$N^>_{\lambda\lambda^\prime}=1+N_{\lambda\lambda^\prime}$.
$M_{\lambda\lambda^{\prime},{\bf q}}$ is the matrix element of
electron-phonon scattering with ${\bf q}={\bf k}_{\lambda}
-{\bf k}_{\lambda^{\prime}}^{\prime}+{\bf k}_{\lambda}^{0}-{\bf k}_{\lambda^{\prime}}^{0}$
being the two-dimensional phonon wave
vector.  The
intra-valley electron--acoustic-phonon scattering is neglected in the present investigation due to the high
temperature. The matrix elements of the intra-valley
electron--longitudinal-optical (LO) phonon scattering in $\Gamma$ valley
and the intra-valley optical-phonon  deformation potential scattering in
$L$ valleys are given by
$M_{\Gamma\Gamma,{\bf
    q}}^{2}=\sum_{q_{z}}\frac{e^{2}\hslash\Omega_{\Gamma\Gamma}}{2\epsilon_{0}(q^{2}+q_{z}^{2})}
    (\kappa_{\infty}^{-1}-\kappa_{0}^{-1})|I_{\Gamma\Gamma}(iq_{z})|^{2}$ and
$M_{L_{i}L_{i},{\bf q}}^{2}=\sum_{q_{z}}
 \frac{\hslash D_{L_{i}L_{i}}^{2}}{2d\Omega_{L_{i}L_{i}}}|I_{L_{i}L_{i}}(iq_{z})|^{2}$, respectively.\cite{lei} For
the  inter-valley electron-phonon scattering, selection rules applied to a cubic zinc-blende structure
of III-V semiconductors show that the LO and longitudinal-acoustic (LA) phonons can assist the inter-valley
transitions.\cite{birman} However, it was further shown later that the selection
rules do not apply in the high energy regime where the inter-valley transfer can happen.\cite{herbert,fawcett,mickevicius}
Therefore, all phonon branches including transverse optical (TO) and transverse acoustic (TA)
phonons can contribute to the inter-valley
scattering.\cite{mickevicius} Nevertheless, the electron-TA phonon scattering is very weak and
can be neglected.\cite{herbert,mickevicius}
The LO, TO and LA phonons have comparable energies,\cite{fawcett} thus the inter-valley
scattering can be considered by grouping together the scattering assisted by these phonons
and using the average phonon energy and the total coupling constant.\cite{mickevicius}
Here the two-dimensional phonon wave vector ${\bf q}$ is mainly determined by the large component ${\bf k}_{\lambda}^{0}-{\bf
  k}_{\lambda^{\prime}}^{0}$, thus the average phonon energy and the total
coupling constant are approximated to be fixed. Therefore, we have
$M_{\Gamma L_{i},{\bf q}}^{2}=M_{L_{i}\Gamma,{\bf
    q}}^{2}=\sum_{q_{z}}\frac{\hslash D_{\Gamma
    L_{i}}^{2}}{2d\Omega_{\Gamma L_{i}}}|I_{\Gamma
  L_{i}}(iq_{z})|^{2}$ for $\Gamma$-$L$ inter-valley electron-phonon
scattering, and $M_{L_{i}L_{j},{\bf q}}^{2}=\sum_{q_{z}}\frac{\hslash
  D_{L_{i}L_{j}}^{2}}{2d\Omega_{L_{i}L_{j}}}|I_{L_{i}L_{j}}(iq_{z})|^{2}$ for $L$-$L$
inter-valley electron-phonon scattering. Here $d=5.36$ g/cm$^{3}$ is the mass density of the crystal;
$\kappa_{0}=12.9$ and $\kappa_{\infty}=10.8$ are the relative static and high-frequency dielectric constants
respectively; $\epsilon_{0}$ is the vacuum dielectric constant.
The phonon energies are $\hslash\Omega_{\Gamma\Gamma}=35.4$\ meV,
$\hslash\Omega_{L_{i}L_{i}}=34.3$\ meV,\cite{lei}
$\hslash\Omega_{\Gamma L_{i}}=\hslash\Omega_{L_{i}\Gamma}=20.8$ meV,\cite{lei}
and $\hslash\Omega_{L_{i}L_{j}}=29.0$\ meV.\cite{lei} The deformation potentials
are $D_{L_{i}L_{i}}$=0.3$\times10^{9}$\ eV/cm,\cite{lei}
$D_{\Gamma L_{i}}=1.1\times10^{9}$\ eV/cm,\cite{lei,mickevicius} and
$D_{L_{i}L_{j}}=1.0\times10^{9}$\ eV/cm.\cite{lei,mickevicius}
$V_{\lambda\lambda^{\prime},{\bf q}}=\sum_{q_{z}}\frac{e^{2}}{\epsilon_{0}
\kappa_{0}(q^{2}+q_{z}^{2}+\kappa^{2})}I_{\lambda\lambda}(iq_{z})I_{\lambda^{\prime}\lambda^{\prime}}^{\ast}
(iq_{z})=\sum_{q_{z}}\frac{e^{2}}{\epsilon_{0}\kappa_{0}(q^{2}+q_{z}^{2}+\kappa^{2})}|I_{\lambda\lambda}(iq_{z})|^{2}$
 is the intra-valley (inter-valley) Coulomb scattering matrix element when
 $\lambda=\lambda^{\prime}$ ($\lambda\neq\lambda^{\prime}$). $U_{\lambda\lambda,{\bf q}}^{2}
 =\sum_{q_{z}}\{Z_{i}e^{2}/[\epsilon_{0}\kappa_{0}(q^{2}+q_{z}^{2}+\kappa^{2})]\}^{2}|I_{\lambda\lambda}(iq_{z})|^{2}$
 is the intra-valley electron-impurity scattering potential, with $Z_{i}=1$ being the charge number
 of the impurity. $N_i$ is the impurity density.
 $\kappa^{2}=Ne^{2}/(a\varepsilon_{0}\kappa_{0}k_{B}T)$ is the
 screening constant. The form factor $|I_{\lambda\lambda^{\prime}}(iq_{z})|^{2}\equiv
 |\langle\phi_{\lambda}(z)|e^{iq_{z}z}|\phi_{\lambda^{\prime}}(z)\rangle|^{2}=\frac{\pi^{4}\sin^{2}y}{y^{2}(y^{2}-\pi^{2})^{2}}$
 with $y\equiv a(q_{z}-K_{\lambda z}^{0}+K_{\lambda^{\prime}
   z}^{0})/2$.

\section{Numerical scheme}
We extend the numerical scheme for solving the KSBEs
at $\Gamma$ valley only, shown in detail in
Ref.\ \onlinecite{weng1}, to the case with both $\Gamma$ and $L$ valleys.
To do so,  five circular zones centered at the bottom of each
valley in momentum space are set up. We divide the truncated two-dimensional momentum space in $\Gamma$ valley into
$N_{\Gamma}\times M$ control regions, and that in
$L_{i}$ ($i$=1, ..., 4) valley into $N_{L}\times M$ control
regions. The relative ${\bf k}_{\lambda}$-grid point in $\lambda$ valley is chosen to be the center of the region,
\begin{equation}
{\bf k}_{\lambda}^{n,m}=\frac{\sqrt{2m_{\lambda}E_{n}^\lambda}}{\hslash}(\cos\theta_{m},\sin\theta_{m}),
\end{equation}
with $E_{n}^\lambda=(n+0.5)\Delta E$ and $\theta_{m}=m\Delta\theta$
($n$=0, 1,$\cdots$, $N_{cut}^\lambda-1$ and $m$=0, 1, $\cdots$, $M-1$). Here
$N_{cut}^\Gamma=N_{\Gamma}$ for $\Gamma$ valley and $N_{cut}^{L_i}=N_{L}$ for $L$
valleys. The truncation energy is $E^\lambda_{N_{cut}^\lambda-1}$. Due to the
effective $\Gamma$-$L$ energy difference,
$E^\Gamma_{N_{\Gamma}-1}-E^L_{N_{L}-1}=\varepsilon_{\Gamma L}$. Thus
$N_{\Gamma}$ and $N_{L}$ are chosen to satisfy
$N_{\Gamma}-N_{L}=[\frac{\varepsilon_{\Gamma L}}{\Delta E}]$
($[\frac{\varepsilon_{\Gamma L}}{\Delta E}]$ is the
integer part of $\frac{\varepsilon_{\Gamma L}}{\Delta E}$). The
energy partition $\Delta E$ is set as $\hbar\Omega_{\Gamma\Gamma}/n_{LO}$
with $n_{LO}$ being an integer, and the angle partition is
$\Delta\theta=2\pi/M$. The coherent terms and the scattering terms of
electron-impurity scattering can be discreted directly. By virtue of
the chosen energy partition, inravalley
electron-phonon scattering in $\Gamma$ valley can be discreted
immediately as well. For intra-valley electron-phonon scattering in $L$ valleys and
inter-valley electron-phonon scattering, one expects to have the
energy difference  be the integer multiplication of
$\Delta E$. This is difficult to satisfy in general. However, with
an optimal value of $n_{LO}$, one can approximately achieve above
requirement. The remaining numerical scheme is all the same as those given in
Ref.\ \onlinecite{weng1}.
\end{appendix}

\end{document}